# Security Assessment of E-Tax Filing Websites

**Aashish Sharma**
NCSA, University of Illinois, Urbana-Champaign
aashish@ncsa.uiuc.edu

**William Yurcik**
NCSA, University of Illinois, Urbana-Champaign
byurcik@ncsa.uiuc.edu

**ABSTRACT**

Technical security is only part of E-Commerce security operations; human usability and security perception play major and sometimes dominating factors. For instance, slick websites with impressive security icons but no real technical security are often perceived by users to be trustworthy (and thus more profitable) than plain vanilla websites that use powerful encryption for transmission and server protection. We study one important type of E-Commerce transaction website, E-Tax Filing, that is exposed to large populations. We assess a large number of international (5), Federal (USA), and state E-Tax filing websites (38) for both technical security protection and human perception of security. As a result of this assessment, we identify security best practices across these E-Tax Filing websites and recommend additional security techniques that have not been found in current use by E-Tax Filing websites.

**INTRODUCTION**

Security is the core of online tax filing system as users are dealing with their highly sensitive financial information. IRS has an aim of getting 80% of the tax returns filed electronically by financial year 2007. Due to this, this is a surge in the number of websites available for filing taxes using Internet. These websites are both hosted by each state's revenue agency as well as private companies trying to earn business. In the process of online tax filing both taxpayer and the state or private tax-filing website have their own security requirements [FROOMKIN]. Typical online tax-filing website requirements are:

- Authentication to identify the taxpayers identity
- Eligibility to insure that the taxpayer possesses required attributes for the tax filing.
- Confirmation to enable the tax website to proof that the taxpayer has authorized the submission of information and filing of the taxes.
- Non-repudiation to protect against taxpayer's denial of submitting false information
- Anonymity to control the amount of information disclosed about a transaction.

Taxpayers also have similar list of security requirements. These requirements include:

- Authentication to confirm the tax website identity
- Integrity to protect against unauthorized access/manipulation of personal financial information
- Confirmation to proof the transaction
- Privacy to control the amount of information disclosed about the transaction to a third party
- Anonymity to control the amount of information disclosed about the transaction to the private third parties assisting the filing of taxes.

Apart from these core security requirements, human factor plays vital role in deciding the usage of these tax-filing websites. If people do not perceive these websites as secure enough they would not use them, even though technically they are running with proper underlying security mechanisms. In this paper, we have tried to identify security and usability features employed by federal, state, and international tax filing websites and exhibit our assessment of security and usability features on these online tax-filing websites.

**RELATED WORK**

(Wenhong, Najdawi 2004) in their study of trust building measures employed by the health related websites showed that health information portals have varying strengthens and weaknesses and may employ only those measures of trust that increases their strength. In addition, these portals may have varying views on the effectiveness of some of the measures and generally, they choose only those which have sufficient influence on enhancing trustworthiness of the websites finally concluding that not all trustworthiness measures could be employed by all portals.

[Whitten A, Tygar J.D., 1999] described the usability issues with respect to security in PGP 5.0 showed that computer security is not usable by the people who are not already knowledgeable in the field. It has showed that usability evaluation criteria often treats security features as primary goals thus creating a wide gap between design perspectives and usage habits where users generally treat security as secondary in using a application.

With e-filing websites we have noticed that designers have tried to carefully place security features and incorporate security mechanisms on the websites, however, the users tend to misinterpret, overlook or perceive these features in a wrong manner. [Princeton Survey Research Associates, 2002] shows users on one hand desire and appreciate obvious security mechanisms on the websites where they wish to transact, but on the other hand subvert these difficult systems and later give up using the application out of desperation.

We think that these problems can be avoided in the online tax-filing websites by a balanced mixture of "visible" security mechanisms and usability features. A perceptually secure website will make users, i.e. taxpayers, feel the "credibility" of the website and create "trust" while ease of use would build confidence in the users to use these websites. [Fogg B.J., 2002] discusses that unlike previous belief people do not use rigorous evaluation schemes but rather focus on design & looks, information structure and content, in the order to access the credibility and trust on online websites. This study emphasizes on the fact that people evaluate credibility of the websites by joining two pieces of information together, one is what they notice about a site and second is the judgments they make on the things they notice. We strongly believe that the users of the online tax-filing websites adhere to this prominence-interpretation theory.

Consumer web watch [Princeton Survey Research Associates (2002)] shows that users have very inconsistent criteria's of accessing websites. They seem to have strict and strong expectations before they arrive to a website and expect the website to provide clear, specific and accurate information about the site policy and practices. However, it has been showed [consumer survey**] that users have strong opinions about information on privacy policy and practices but it does not mean that these users are aggressive in seeking this information.

Efforts have also been made in developing community/group efforts for accessing websites by suggesting to create reliable reputation reporting mechanisms for online communities [Dellarocas, C., 2001]. [Turner C, Zavod M, and Yurcik W, 2001] have also discussed factors that affect perceptions of security and privacy of E-Commerce websites in their study.
In effort to understand net users attitudes about online privacy and security AT&T has also conducted a study [AT&T Labs-Research Technical Report, TR 99.4.3] emphasizing of user's viewpoint about categorization of some data to be of more importance then other. It has been found out via this study that users are less reluctant to provide their postal address then their active email address then their phone numbers. This report shows that according to the purpose of the application the behaviour of the user changes towards the security and sensitivity of the application for example users attitude for accepting cookies changes with the type of websites they are visiting or they tend to dislike automatic data transfers on a website.

**TYPICAL SECURITY AND USABILITY FEATURES ON TAX-FILING WEBSITES**

A tax filing website may have the following security and usability features: authentication mechanism, Security connections using SSL, privacy policies, third party trust symbols, post tax filing procedure, application usage and eligibility criteria, demos and FAQ's.

Generally, all the websites were found to have most of the above security and usability features with some features emphasized heavily upon by some websites, mainly, depending on the hosting agency (state or private companies). We found that private companies were emphasizing heavily on trusted third party symbols where as state tax websites were using their advantage of access to the taxpayer's data by facilitating authentication using social security numbers and a pin supplied by them.

In the following sections, we briefly state the application of these features mostly found on state taxa filing websites.

**STATE TAX FILING WEBSITES**

In total 38 states facilitate online-tax filing. Some of the states have, like IRS, collaborated with the various private tax filing website while others have their own portal for Internet tax filing for their residents. Apart from the tax filing, Internet is also significantly used for tax administrative functions. Most taxing authorities now provide copies of their forms, in downloadable format, on their websites. Many offer online inquiry into the status of individual income tax refunds. The more advanced administrative uses of the Internet, however, center on the areas of account maintenance and customer service.

Account maintenance includes taxpayer registration, name and address changes, the ability to check account balances and outstanding liabilities, and the ability to pay outstanding balances online. The movement is toward self-maintenance, that is, the taxpayer actively maintains the information in his own account, rather than having to provide the information to tax authority personnel. Self-maintenance increases accuracy, by eliminating third-party keying of data.

In addition to general information, these websites are beginning to maintain Frequently Asked Questions (FAQ's) on a variety of tax topics in a format that is simple for the customer to navigate. Best practices also include linking the website to the taxing authority's electronic mail system, so that the customer can enter e-mail directly from the website, targeted to the specific area capable of responding to a particular subject area.

De facto standards for the security of interactive web-based applications include the use of PINs or passwords, and the use of secured socket layer (SSL) so that the sender and the intended receiver can only read the transmission. PIN's are issued to taxpayers by an independent method (usually paper mail) and/or are verified against a database by the Internet application.

**Security Connections**

SSL is deployed on all the websites (state and private) for the secure communication and is highlighted reasonably well on all the sites assuring the users of security.

**Authentication mechanism**

In order to grant access to the genuine user for filing her/his tax returns all of the websites are employing authorization mechanisms. Different tax websites have employed various different ways to identify the tax filers primary being social security number. However, we have also seen states employing users to fulfil other formalities like submission of electronic funds transfer from (EFT) (Alabama) where a user can choose her/his username of choice by mentioning it in the form. Arkansas web filing system required users to supply a pin printed on their tele-file pamphlet that taxpayers in Arkansas receive in mail; along with Primary taxpayers and secondary taxpayers social security numbers.

Colorado requires users to use SSN, a pre-supplied pin number along with Refund amount and estimated amount of tax due this year and last five digits of billing number. Its highly likely that users might not be able to have everything in hand so the state website provides you with a facility to request all this in postal mail.

Illinois requires IL Pin to be used along with social security number but facilitates an online delivery on IL pin based on requirements that you are able to supply your driver's license number or estimated tax refund you are getting or how much tax have you paid last year.

Private companies such as HR Block, Tax Brain amongst others ask users to create their own custom login names and password (email address and a 6-12 alpha-numeric password) for authentication purposes.

**Third party trust symbols**

Third party trust symbols have been more prevalent in private tax filing websites as compared to state websites. The most logical explanation could be that state revenue websites do carry credibility of their own.

These third party trust symbols have not been real security gauge on the websites for various reasons. The foremost reason is that most of these websites have placed these symbols along with advertisements or animated gif's images due to which generally users tend to ignore them. We have also found misuse of third party symbols in websites where some of the signs have been found to have no verification links but just graphic images.

IRS has also issued its own e-file picture and has encouraged the various partners in free-file-alliance to use this graphic. This graphic also lacks verification system and anyone with an expertise of putting images on the web pages can put this symbol of their website and there is virtually no way to determine if IRS as approved usage of the symbol on that particular website.

**Privacy Policies**

[Fogg B.J., 2002] in their study also noted that users tend to aggressively look for a privacy policy on the website however, this does not necessarily mean that users also read them. All of the states and private tax websites have very well implemented this. Privacy and security policies are visible on all the websites and sometimes the links are available on multiple pages for users to be able to easily find these policies.

We have found some discrepancies in the system of privacy policies. In certain tax filing websites, multiple sub companies are handling tax-data collection and tax data submission and help desk responsibilities with each having their own privacy policies. Users need to aware of this and should cautiously see if a particular tax website has multiple privacy policies.

**Post tax-filing procedures**

It is mandatory by law for the IRS and other tax filing websites to acknowledge the receipt of the returns filed by the taxpayer. If the taxpayer does not get the receipt, taxes are not officially accepted as filed. Here, we recommend the standard way of acknowledging the taxpayer via email is a good thought. An email with appropriate information about the receipt of the tax data as well as links to important questions, along with telephone numbers and contact information would only gain taxpayers confidence in the system. By no means this email should be very flashy or give feel of a spam to the users.

**Application usage and eligibility criteria**

Tax filing system in the United States is one of the most complex systems and hence its pretty difficult to have a universal one stop shop portal to facilitate the tax filing. Eligibility criteria, thus, becomes very important for the users. No taxpayer would appreciate that after s/he has filed half of the data, application notifies that they are not eligible to file taxes using this website. We found that eligibility criteria has not always been placed at visible locations, with some websites putting in FAQ's section assuming that everyone accessing the website is reading FAQ's.

**Demo Present**

It is very much possible for humans to loose sense of direction or location in a hypertext environment due to additional effort and concentration necessary to maintain several tasks simultaneously, cognitive overload. Demo is a very important part of tax filing websites. Since tax-filing application is a totally new domain for most of the users they are not very familiar with what exactly is this application. To add to this, first time users have a general curiosity about how exactly the system work. People would want to find out first before using this application.

The advantage if demo is that it helps building mental model for user about the application. It helps users to create a mental navigation map and gives ability to keep conscious track of the links. On seeing a demo a user gets fairly well acquainted with the system and is aware of what to expect when actually using the online tax website to file taxes.

**FAQ's available**

FAQ's also from an integrated requirement of a tax filing website. Users, especially first time users, have lots of questions and curies about the system and the entire process of how exactly it works. FAQ's on the website can really help bolster confidence of a new or a curious user by providing the desired information to him/her. Also another advantage of this is it reduces a vast amount of burden from customer support by reducing support calls.

We have observed that some websites have very detailed FAQ's section (Indiana) however, the entire FAQ section is one very huge html file with no subsections or categorization of questions and answers and virtually no means to search information in the file. This particular example illustrates how faq's can at time do more harm then good.

**FEDERAL TAX FILING WEBSITES**

IRS does not have its own federal tax website, instead it uses third party alliance to assist the taxpayers in filing federal taxes via Internet. To facilitate this IRS has started a program called E-File. All the private companies which want to provide services to the residents need to strictly comply to IRS's regulation for E-Filing and those companies which passes these regulations are listed on IRS's website as e-filing partners. IRS does not take any responsibility for these websites but it guarantees that these websites have met online filing criteria decided by IRS.

The key feature of these private websites is their commercial aspect. These website put advertisements, third party symbols, promotional offers in form of free filing (if you qualify) etc in order to gain business. We surveyed a some of the available websites for security and usability criteria mentioned in [Fogg 2002]. The following table shows our survey of these private websites.

| Feature | Free1040taxreturn | Taxcite.com | Quicken | Esmarttax |
|---|---|---|---|---|
| Does this site has a demo for tax | Yes | Yes | Yes | Yes |
| Does the site uses the electronic version of form | Yes | Yes | Yes | |
| Does the site provides information about specific fields in forms | Yes | Yes | Yes | |
| Does the site has any FAQ section | No | Yes | Yes | Yes |
| Does the site tells anything about its functionality of how it process the taxes | No | No | Yes | No |
| Does the site talks about what security they are using | Yes 128 bit encryption | Yes | Yes | Yes |
| Does the site talk about the browser compatibility | Yes | No | Yes | No |
| Does the site gives links for download of latest browsers | No | No | No | No |
| Does the site uses https | Yes | Yes | Yes | Yes |
| Does the site has promotional offers | No | Yes | Yes | no |
| Does the site sells anything else then tax filing services | No | Yes books mostly related with tax. Links with Amazon.com | No | No |
| Does the site provides more then personal tax filing example - corporate accounts management etc | No | Yes | No | Yes |
| Is the site down after taxes are filed | Yes | Bad links half functionality that too improper | Customized | No |
| Does the site has some trustworthy symbols | Yes not visible easily. May be sister website | Yes. Not clear though | No. (brand Image) | Yes. IRS authorized banner |
| Does the website talks about what if there is error in the forms or filling | No | No | No | Yes |
| The site provides a quick response to your customer service questions. | | | | |
| The site lists the organization's physical address. | Yes | Yes | | Yes |
| The site gives a contact phone number. | Yes | Yes | No | Yes |
| The site gives a contact email address. | Yes | Yes | No | Yes |
| The site shows photos of the organization's members. | No | | No | |
| | | | | |
| The site lets you search past content (i.e. archives). | bad navigation. Found cyclic links. | No | Yes | Yes |
| The site looks professionally designed.(on scale of 1-10) | 2 | 3 | 5 | 5 |
| The site is arranged in a way that makes sense to you. | Some what | some what | Yes | Yes |
| The site takes a long time to download. | 40 Sec. At 28.8 kbps | 33 sec on 28.8 | 30s on 28.8 | 44 sec on 28.8 |
| The site is difficult to navigate. | One cyclic link | No | No | No |
| A site you think is believable links to the site. | No | Yes | Yes | Yes |
| The site states its policy on content. | Amateurish | No | Yes | Yes |
| The site recognizes that you have been there before. | Yes | Yes | No | Y (option if the user wants to) |
| The site automatically pops up new windows with ads. | No | No | No | No |
| The site makes it hard to distinguish ads from content. | No | Yes | No | No |
| The site has been updated since last visit. | No | | No | Yes |
| The site offers information in more than one language. | No | No | No | No |
| The site is small (e.g. less than 5 pages). | Yes | No | No | No |
| The site is hosted by a third party (e.g. AOL, Geocities). | No | No | No | No |
| The site's domain name Does not match the company's name. | Yes | No | No | Yes |

**INTERNATIONAL TAX-FILING WEBSITES**

In order to examine how the tax filing systems of other countries work we did a comparative study of online tax-filing system of 5 different countries; Australia, Britain, Canada, Singapore and The United State of America.

We evaluated these websites on the criteria's of affiliation. Canada, United States, and Australian tax authorities use third parties (private companies) to assist tax payer in filing taxes online. Third parties have to qualify the guidelines of the respective tax authorities to be able to enroll in the e-tax program. However, Britain and Singapore do not use any third parties but rather have the government run the e-tax filing websites.

The following table shows a comparative analysis of the tax filing system of these countries.

| Properties | Australia | Britain | Canada | Singapore | United States |
|---|---|---|---|---|---|
| Affiliation | Australian Taxation office | Office of Inland Revenue | Canada Revenue Agency | Inland Revenue Authority of Singapore | US Treasury Department |
| Authentication Mechanism | Use software to do taxes and post the printout to the tax authority. | Digital certificates provided to individuals<br><br>Login/password obtained via registration | Authentication of Social Insurance Number and pin supplied by agency on paper | Tax Payer No and EF Pin both issued by govt on paper for taxpayer SingPass | Varying. Uses SSN (State), email + password (private) |
| Availability | July onwards | 24/ 7. Use of same website for other government purposes. | 21 Hours/day (Feb 4 – Sept 30)/Y | Only in tax filing season February 20th onwards | 24 / 7<br><br>Filing season |
| Eligibility Criteria Mentioned | Yes | Yes | Yes | Yes | Yes |
| Security / Privacy Policies/Legal Notices | Yes | Yes | Yes | Yes | Yes |
| SSL Connection | Information posted | Uses digital certificates given to individuals | Yes | Yes | Yes |
| Third Party services needed | Yes. Software from third parties is needed to generate paperwork for posting to tax authority. | None | Yes, Netfile is just a transmission service. Need to install software certified by the netfile to create file and send it. | None | Yes. IRS defines guidelines to facilitates state and third party for tax filing |
| Third Party Seals/Symbols | None | No | No |  | Own E-File Symbol |
| Website URL | www.ato.gov.au | https://online.inlandrevenue.gov.uk | www.netfile.gc.ca | www.iras.gov.sg | www.irs.gov |

**Table-1. Comparison of International Tax Internet Tax filing system**

**Conclusion**

Usability studies have showed that users become more confident about the security of the system when they see visible security. There is no visual feed back to the user about connection establishment of secure connection by their browser. Usually a secured website is designed in compartments of public accessible sections and non-public accessible sections with authentication required. If users are able to see these compartments on their webpage, it will definitely bolster their confidence on the websites security. A visual feed back of showing secure zone transition would definitely help users build mental model for establishing secure connection.

We feel that the users would become more aware about security and https protocol then relying on the usual padlock sign () at the bottom of the browser's window that is greatly gone unnoticed. Our design in both secure section and general public section is consistent so that user is aware of the look and feel and functionality of the website.

Adding Reverse Turing Test with authentication system in order to deter automated brute force attacks on the websites would also be a good visible security symbol. Reverse Turing Test requires the users to start typing words or phrase embedded in a graphic after supplying a wrong password a given number of times.

This mechanism is employed by lots of e-commerce websites like paypal and yahoo mail. However, this system needs to be full proof to the attacks. For example, we generate random filenames for the graphic we supply so that attackers cannot identify a particular graphic and associate it with the word embedded into it.

We strongly feel that the state and private tax-filing websites should adopt Reverse Turing Test as a security mechanism to stop automated brute force attacks on the websites after which the social security number and supplied pin may become irrelevant.

State websites have refrained from using third party symbols, which, to our belief, would not affect a visiting taxpayers decision about security.

We feel that FAQ's section is an ingredient part of a tax website (like address book is for an email application) however, they should be designed based on topical relevance. A proper categorization of information (questions and answers) in faq's along with a proper easy to use but effective search mechanism can make them a very usable tool in tax-filing websites. Also, the link for faq's should be place in a highly visible section of home page.

Faq's are good idea but to create user community on online tax websites especially for states would enhance the credibility of the system a lot. These can function as moderated new groups where experts from support desk can answer questions if the community is unable to figure out a solution to a genuine concern.

It is important that you write for the general audience. It is also important that your work is presented in a professional fashion, which is what this guideline is intended to help you with. By adhering to the guideline, you also help the conference organizers tremendously in reducing our workload and ensuring impressive presentation of your conference paper. We thank you very much for your cooperation and look forward to receiving your nice looking, camera-ready version!

**SELECTION OF WEBSITES**

For the purposes of brevity, we chose websites that optimally portrayed security and usability features discussed in this paper. We found tremendous consistency in the websites designed in particular fashion. For example, the states that do not have their own web filing system and have collaborated with third parties have very similar design features (example Alabama, Alaska, Arizona, Arkansas, Georgia, Hawaii, Idaho, Rhode Island, Oregon, North Dakota, New Mexico, New Hampshire etc). Like wise, we found that state having own portal for direct tax filing (example Illinois, Connecticut, Indiana, Maryland, New Jersey etc) have similar design and authentication system.